# Predicting input impedance and efficiency of graphene reconfigurable dipoles using a simple circuit model

Michele Tamagnone, *Student member, IEEE*, Julien Perruisseau-Carrier, *Senior Member, IEEE*

*Abstract*— An analytical circuit model able to predict the input impedance of reconfigurable graphene plasmonic dipoles is presented. A suitable definition of plasmonic characteristic impedance, employing natural currents, is used to for consistent modeling of the antenna-load connection in the circuit. In its purely analytical form, the model shows good agreement with full-wave simulations, and explains the remarkable tuning properties of graphene antennas. Furthermore, using a *single* full-wave simulation and scaling laws, additional parasitic elements can be determined for a vast parametric space, leading to very accurate modeling. Finally, we also show that the modeling approach allows fair estimation of radiation efficiency as well. The approach also applies to thin plasmonic antennas realized using noble metals or semiconductors.

## 1. Introduction

THE ELECTROMAGNETIC properties of graphene for microwave, terahertz and infrared applications have been explored in numerous works [1-13]. In particular, graphene dipole antennas based on plasmonic modes have been recently proposed for terahertz and far infrared frequencies [1-5]. The device in [1] consists of two graphene patches placed on a substrate and separated by a small gap, which hosts the lumped element – source or detector – connected to the antenna. The antenna high input impedance at resonance provides a good matching to lumped terahertz sources and detectors (e.g. photomixers and thin film diodes) which are generally characterized by very high impedance values [1]. Furthermore, the resonance of the antenna can be dynamically tuned in frequency using the electric field effect, while preserving the antenna impedance and radiation characteristics [1, 3].

When no lumped element is connected to the antenna, a strongly enhanced electric field is created in the gap at the resonance frequency. The antenna can then behave as a scatterer whose properties strongly depend on materials in proximity of the gap, enabling sensing application at THz and infrared frequencies [3].

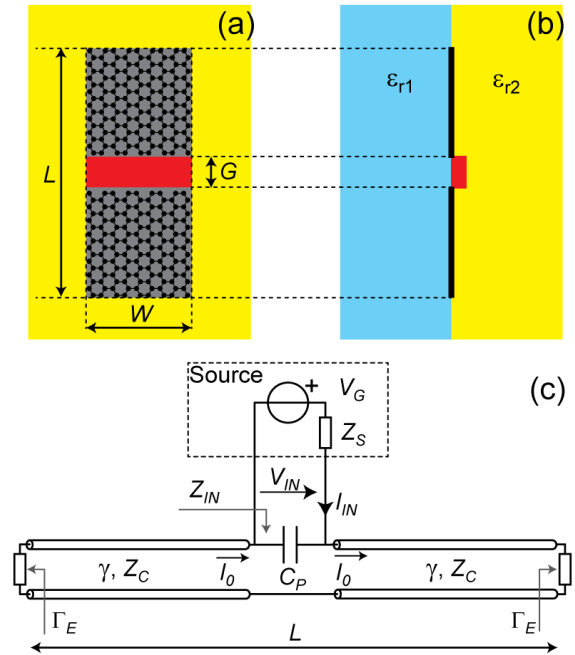

Fig. 1. Graphene plasmonic dipole antenna and TL model. (a) Top view and geometrical parameters. (b) Lateral view showing the superstrate and substrate permittivities. (c) Complete transmission line model of the plasmonic antenna.

The properties of the proposed graphene dipoles are strongly dependent on geometrical dimensions, on graphene highly-variable parameters, and on the substrate. Consequently, the optimization of the antenna involves a search in a vast parameter space, even for very simple geometries.

To facilitate the design of such antennas as well as their understanding, here we provide an analytical circuit model for rectangular tunable graphene dipoles, which can accurately predict the input impedance properties upon sweeps in the parameter space. The model can be used to significantly improve the time required for the optimization with respect to purely full-wave approaches. Furthermore, it can be easily extended to similar plasmonic antennas based on noble metals

This work was supported by the Hasler Foundation (Project 11149) and by the Swiss National Science Foundation (SNSF) under grant 133583.

The authors are with the AdaptiveMicroNanoWave Systems Group, LEMA/Nanolab, Ecole Polytechnique Fédérale de Lausanne, 1015 Lausanne, Switzerland (e-mail: michele.tamagnone@epfl.ch; julien.perruisseau-carrier@epfl.ch).

at optical frequencies.

## *2. Circuit model*

Figure 1 shows the antenna modeled in this work. The graphene patches are rectangular and are placed between a superstrate (with permittivity $\varepsilon_{r1}$) and a substrate (with permittivity $\varepsilon_{r2}$). The source/detector is placed in the small gap $G$ and it is connected to the two patches. The inductive conductivity of the graphene patches at THz frequencies allows the propagation of TM plasmonic modes which can be excited by the source [1].

The first step towards the antenna model consists in determining the equivalent transmission line (TL) parameters (namely the propagation constant $\gamma$ and the characteristic impedance $Z_C$) for a plasmonic TM mode on a graphene strip of width $W$. For sufficiently large $W$, the propagation constant is very similar to the one of plasmons propagating on an infinitely wide plasmonic sheet. The latter can be expressed in a well-known closed form solving Maxwell equations above and below the graphene sheet and using the impedance of graphene as boundary condition [8]. This procedure yields the following dispersion equation, which can be solved for $\gamma$:

$$\frac{\varepsilon_{r1}}{\sqrt{-\gamma^2 - \varepsilon_{r1}k_0^2}} + \frac{\varepsilon_{r2}}{\sqrt{-\gamma^2 - \varepsilon_{r2}k_0^2}} = \frac{j\sigma}{\omega\varepsilon_0} \quad (1)$$

where $\sigma$ is the surface conductivity of graphene (doubled in case two stacked sheets are used as in [1]), $\omega = 2\pi f$ is the frequency, $\varepsilon_0$ is the vacuum permittivity, $k_0 = \omega/c$.

Though the definition of a characteristic impedance is in general arbitrary, two conditions can be enforced to ensure a consistent modeling of the current in the source in the model of Fig. 1:

- Condition (i): the current on the equivalent TL must be equal to the total current on the graphene strip (in other words the current in the equivalent TL model is the so-called 'natural current').
- Condition (ii): consequently, the loss is modeled by a resistance in series with the inductive part of the TL equivalent LC cell.

The TM plasmonic mode on an infinite graphene sheet is characterized by a power density $P_{sheet}$ and by a surface current density $J$ on graphene. The total current on the strip is approximated by $I = WJ$, and similarly the total power on the strip is $P_{strip} = WP_{sheet}$. $P_{sheet}$ can be found by integrating the Poynting vector on the axis $x$ perpendicular to graphene and to the propagation direction:

$$P_{sheet} = \hat{\mathbf{z}} \cdot Re\left(\int_0^{+\infty} \mathbf{E}_1 \times \mathbf{H}_1^* \, dx + \int_{-\infty}^0 \mathbf{E}_2 \times \mathbf{H}_2^* \, dx\right) \quad (2)$$

where $\hat{\mathbf{z}}$ is the propagation direction. The integral in (2) is split in two parts: the first term for the evanescent fields $\mathbf{E}_1, \mathbf{H}_1$ in the superstrate, and the second for the evanescent fields $\mathbf{E}_2, \mathbf{H}_2$ in the substrate. Since the fields and currents associated to the propagating TM mode are known (not reproduced here but available in [8]), equation (2) allows to analytically express $P_{sheet}$ as a function of $J$:

$$P_{sheet} = \frac{Im(\gamma)\omega\varepsilon_0\varepsilon_{r1}|J|^2}{2|\sigma|^2|\gamma_{x1}|^2 Re(-\gamma_{x1})} + \frac{Im(\gamma)\omega\varepsilon_0\varepsilon_{r2}|J|^2}{2|\sigma|^2|\gamma_{x2}|^2 Re(\gamma_{x2})} \quad (3)$$

where $\gamma_{x1,2} = \mp j\sqrt{(\gamma^2 + \omega^2\mu_0\varepsilon_0\varepsilon_{r1,2})}$.

Consequently, $P_{strip}$ can be expressed as function of $I$, and the real part of the characteristic impedance $Z_C$ of the TL model can be determined using condition (i) above as:

$$Re(Z_C) = \frac{Im(\gamma)\omega}{2W|\sigma|^2}\left(\frac{\varepsilon_0\varepsilon_{r1}}{|\gamma_{x1}|^2 Re(-\gamma_{x1})} + \frac{\varepsilon_0\varepsilon_{r2}}{|\gamma_{x2}|^2 Re(\gamma_{x2})}\right) \quad (4)$$

The imaginary part of the impedance can be found using condition (ii), which requires $Re(\gamma/Z_C) = 0$. Hence:

$$Im(Z_C) = -\frac{Re(\gamma)}{Im(\gamma)} Re(Z_C) \quad (5)$$

Figure 1c shows the overall circuit model which uses the derived transmission line parameters. The source is connected to two transmission line stubs which model the two patches of the antenna. Each patch can be seen as a TL section approximately terminated by an open circuit, since the natural current vanishes at the antenna ends. However, the presence of fringing fields (an example is depicted in figure 2) results in a reflection coefficient $\Gamma_E$ different from 1 and thus a load is introduced as a termination of the transmission line model. Parasitic fields are found in the gap as well and can be included in the model adding a capacitance $C_P$ in parallel with the source.

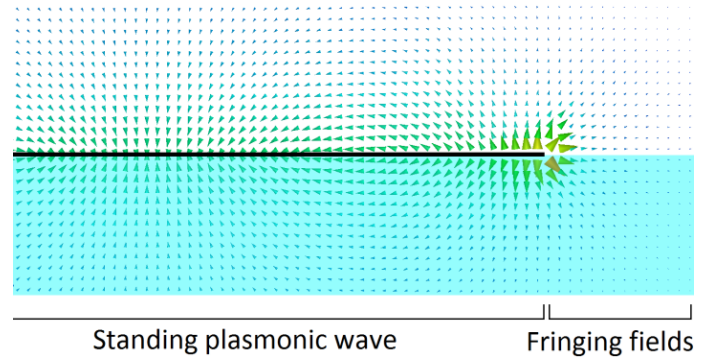

Fig. 2. Geometry of the fringing electrical field at the antenna edge. A large value of $L$ has been used to better show the standing plasmonic wave caused be the reflection at the edge.

Several important considerations on these parasitic elements can be made. Firstly, it is straightforward to see that the electrostatic coupling between the two half patches edges is proportional to the width of the strips. Hence the ratio $C_P/W$ is expected to be constant for different geometrical parameters. Furthermore the reflection coefficient $\Gamma_E$ is approximately independent from $W$, $L$, frequency, and propagation constant. In fact it can be shown that, in the high confinement limit discussed above, the fringing fields and the plasmonic modes undergo a simple geometrical scaling if the frequency or the propagation constant is changed. Since $\Gamma_E$ is defined as the ratio between incoming and reflected wave, its value is not affected by the scaling. Consequently, this approach is more accurate than ones based on equivalent length (as used in [5] to predict the resonant frequency of rectangular scatterers).

The fact that the TL model is based on 'natural currents' has several advantages. For example, it enables the estimation of the radiation efficiency $\eta_{rad} = P_{rad}/P_{in}$ of the antenna, where $P_{rad}$ is the radiated power and $P_{in}$ is the total power accepted at the antenna port. This can be achieved by determining the current distribution on the TL model due to an arbitrary source (defined by $V_G$ and $Z_S$) and then applying directly the radiation integral. The total radiated power $P_{rad}$ is then found integrating for all directions. The antenna shown in [1], as most plasmonic antennas, is electrically small. For this limit, $P_{rad}$ admits a close form expression obtained approximating the radiation pattern to the one of an Hertzian dipole:

$$P_{rad} = \frac{8\pi\eta_0 |I_0|^2}{3\lambda^2 |\gamma|^2} \cdot \frac{\sqrt{\varepsilon_{r1}} + \sqrt{\varepsilon_{r2}}}{2} \cdot \frac{\left(1 - e^{-\frac{\gamma L}{2}}\right)\left(1 - \Gamma_E e^{-\frac{\gamma L}{2}}\right)}{(1 - \Gamma_E e^{-\gamma L})} \quad (6)$$

where $\eta_0$ and $\lambda$ are the free space impedance and wavelength respectively, and $I_0$ is the current as defined in Figure 1c. Concerning $P_{in}$, it is easily determined as $P_{in} = V_{in} I_{in}^*$, where the input voltage and current $V_{in}$ and $I_{in}$ can be computed since all the circuit parameters are known. Similarly, the total efficiency is given by $\eta_{tot} = P_{rad}/P_{av}$, where $P_{av}$ is the source available power. Of course, unlike $\eta_{rad}$, $\eta_{tot}$ depends also on the source internal impedance $Z_S$.

Equations (4) and (6) show that the radiation efficiency $\eta_{rad}$ of the antenna is proportional to the width $W$. In fact, for a given current, $P_{rad}$ is independent of $W$ according to (6), while the input power is decreased for larger $W$, as evident from (4). This fact motivates the large $W$ limit, for which better designs can be achieved.

### *3. Results*

The proposed model has been validated against full wave simulations performed using the commercial software Ansys HFSS. The bidimensional complex frequency-dependent conductivity of graphene has been computed using the Kubo formalism and implemented in the simulation using the 'impedance' boundary condition provided by the software [1]. Figure 3a shows the comparison between the input impedance computed via full-wave simulation and that predicted by the model under the assumption that the effects of the parasitic quantities $\Gamma_E$ and $C_P$ are negligible, i.e. $C_P = 0$ and $\Gamma_E = 1$ (perfect open circuit). The results show that the model successfully predicts the input impedance stability upon reconfiguration observed and discussed in [1]. The maximum real input impedance and the resonance frequency are both predicted with relative errors around 20%. Hence, even for unknown $\Gamma_E$ and $C_P$, the model can be used for an initial rough design of the antenna.

However for optimization purposes a higher precision is necessary. This can be achieved performing a *single* full-wave simulation on the initial design and then fitting the model to it by fine-tuning $C_P$ and $\Gamma_E$. Then, the behavior of $C_P$ and $\Gamma_E$ described in section II allows a fine optimization using the model with constant values of $C_P/W$ and $\Gamma_E$. Figure 3b shows that the model now accurately predicts the input impedance. If graphene parameters are changed (the figure shows a $\mu_C$ sweep as an example) the accuracy is preserved. Furthermore the frequency of the second resonance of the antenna is also in excellent agreement, providing strong evidence that the dispersion relation is in agreement with the one expected from the infinite sheet theory.

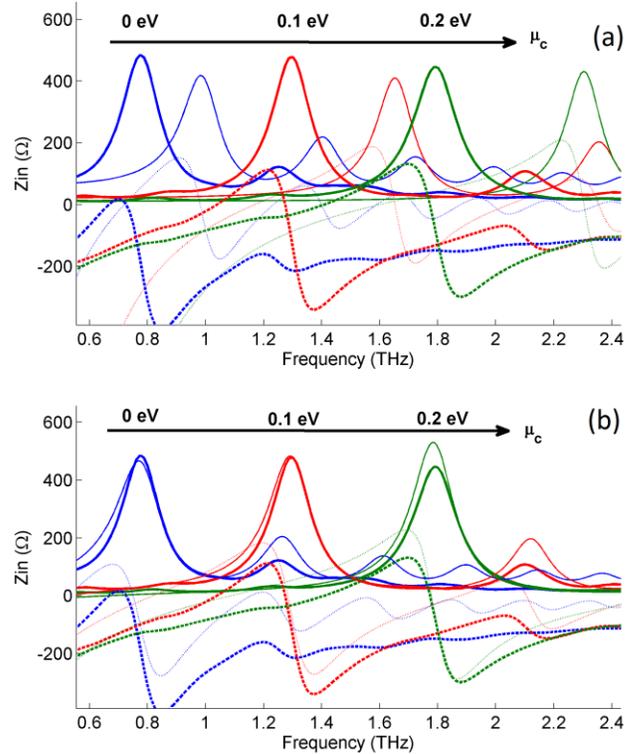

Fig. 3. Comparison between input impedance using full-wave simulations (thick traces) and the proposed TL model (thin traces). Dashed traces represent the imaginary part of the impedance and solid ones the real part. Geometrical and graphene parameters are the same used in [1]. (a) Model with $\Gamma_E = 1$ and $C_P = 0$. (b) Model with $\Gamma_E = e^{-j1.57}$ and $C_P = 0.35$ fF.

Figure 4 compares the total efficiency predicted by the model through equation (6) and the one obtained using full wave simulations. The agreement is excellent for low frequencies, while discrepancies are observed above 1.3 THz. These differences are probably due to higher order plasmonic modes which do not contribute to radiation and to minor discrepancies observed in the input impedance in figure 3b.

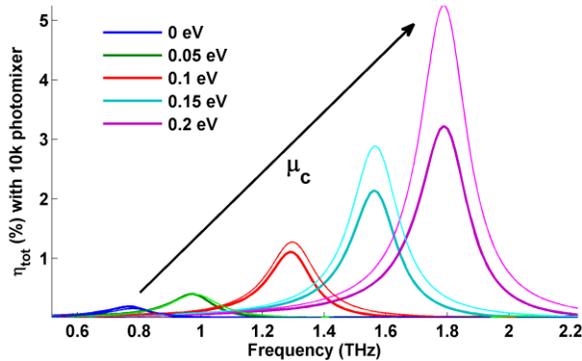

Fig. 4. Comparison between total efficiency using full-wave simulations (thick traces) and the proposed TL model (thin traces). The considered source is a THz photomixer having internal impedance of 10 kΩ. Geometrical and graphene parameters are the same used in [1]. Model with $\Gamma_E = e^{-j1.57}$ and $C_P = 0.35$ fF.

## *4. Conclusion*

A model for reconfigurable graphene plasmonic dipoles has been proposed, which allows performing fast optimizations of graphene dipole antennas considering both input impedance and efficiencies.